\documentclass{article}

\usepackage[preprint,nonatbib]{aaj2026}
\usepackage[numbers]{natbib}
\usepackage[utf8]{inputenc}
\usepackage[T1]{fontenc}
\usepackage{hyperref}
\usepackage{url}
\usepackage{array}
\usepackage{booktabs}
\usepackage{amsfonts}
\usepackage{nicefrac}
\usepackage{microtype}
\usepackage{xcolor}
\usepackage{caption}
\usepackage{tikz}
\usetikzlibrary{fit,positioning,backgrounds}
\definecolor{ctreq}{HTML}{EEDCDC}
\definecolor{ccont}{HTML}{DCE3EE}
\definecolor{cdesign}{HTML}{DCEEDC}

\title{Cognitive Comparability and the Limits of Governance:\\[0.3em] \large Evaluating Authority Under Radical Capability Asymmetry}

\author{%
  Tony Rost\thanks{ORCID: \href{https://orcid.org/0009-0008-0637-6654}{0009-0008-0637-6654}} \\
  Portland, OR, United States \\
  \texttt{tony@sigov.institute} \\
}

\begin{document}

\maketitle

\begin{abstract}
Governance theory presupposes a rough cognitive comparability between governors and governed. This paper makes that assumption explicit and testable through a six-dimension evaluation framework covering legitimacy, accountability, corrigibility, non-domination, subsidiarity, and institutional resilience, drawn from political legitimacy theory, principal-agent models, republican theory, and the AI alignment literature. The framework is first demonstrated on existing non-majoritarian institutions, where capability asymmetry is real but bounded, and then applied to a prospective case of bounded superintelligent authority, where the asymmetry is radical. Four of six dimensions show structural failures. Two of the four appear tractable to institutional design (subsidiarity scope limitation and institutional resilience). The other two, the public reason problem under cognitive incomprehensibility and the non-domination problem under permanent capability asymmetry, call for new normative theory rather than better institutional design. The analysis also finds that dimensions which operate as independent checks under bounded asymmetry begin to degrade together under radical asymmetry, because each depends on the same oversight capacity. The assumptions that allowed these checks to remain independent have gone unexamined so far because they have always held.
\end{abstract}

\noindent\textbf{Keywords}: artificial superintelligence; governance theory; cognitive comparability; capability asymmetry; political legitimacy; principal-agent theory; non-domination; multi-level governance; AI alignment; institutional design; public reason

\section{Introduction}
\label{sec:intro}

Repeated surveys of AI researchers show a sustained expectation that AI systems may match or surpass human-level performance across a broad range of cognitive tasks within the coming decades. \citet{grace2018will} found that median estimates for a 50\% probability of high-level machine intelligence varied substantially with question framing, ranging from the mid-2050s to the late 2060s, while follow-up surveys have shown these estimates shifting earlier still. A 2023 survey of 2,778 researchers reported a median estimate of 2047~\cite{grace2025thousands}. These forecasts carry substantial uncertainty and documented methodological limitations, including the framing sensitivity just noted and regional variation. They do not establish that artificial superintelligence will be developed, but they do establish that a significant fraction of AI researchers take the possibility seriously. Combined with assessments of existential risk from advanced AI~\cite{carlsmith2022power, bostrom2013existential, ord2020precipice}, they give governance theory reason to develop evaluative tools now, before specific proposals become urgent.

The question has gained practical urgency as leaders of major AI laboratories project artificial superintelligence within years, while technology leaders and AI researchers have begun framing superintelligence development as a geopolitical challenge requiring national security strategy~\cite{hendrycks2025superintelligence}. The Future of Life Institute's statement on superintelligence has gathered over 133,000 signatories, including leading AI researchers, calling for international action to prevent uncontrolled development~\cite{fli2025statement}. The United Kingdom's House of Lords debated a superintelligence moratorium in early 2026, and prominent researchers have argued that ASI development without explicit humanity-wide consent constitutes an impermissible imposition~\cite{aguirre2025keep, yudkowsky2025anyone}. Prevention of unsafe ASI development is one response to these concerns; governance frameworks for the case where prevention is incomplete or unsuccessful are another. The two are not mutually exclusive: responsible preparation requires both. This paper contributes to the second.

A large and growing literature addresses the governance \textit{of} AI~\cite{dafoe2018ai, floridi2019unified, calo2017artificial, erdelyi2018regulating, roberts2024global}, now surveyed comprehensively in \citet{bullock2024oxford}. A separate technical literature addresses the alignment problem: how to ensure that advanced AI systems pursue objectives consistent with human values~\cite{russell2019human, amodei2016concrete, ngo2022alignment}. Recent work has begun to address the political legitimacy of AI in governance roles~\cite{erman2024artificial}, while \citet{han2025question} taxonomizes AI governance into four categories along two axes (democratic versus non-democratic process, and binding versus non-binding authority), arguing that the most pressing governance challenges arise in non-democratic binding arrangements. The bounded superintelligent authority examined here falls squarely in Han's most problematic quadrant: binding authority without democratic process. The framework below deepens Han's taxonomy by specifying conditions under which such authority might satisfy governance requirements despite its non-democratic character; radical capability asymmetry precludes satisfaction on multiple dimensions.

\citet{bostrom2020public} come closest to the present inquiry, identifying policy desiderata for superintelligent AI through a ``vector field'' approach that maps the direction various structural features of the scenario push policy, regardless of normative starting point. Their analysis establishes that superintelligence governance requires distinctive policy considerations but does not specify necessary conditions or construct a failure taxonomy. The framework extends that desiderata approach: specifying dimensions, identifying necessary conditions along each, and classifying failures as contingent, design-tractable, or theory-requiring. Other relevant discussions appear in industry proposals~\cite{altman2023governance}, technical AI safety work~\cite{bostrom2014superintelligence, russell2019human}, or algorithmic governance literature~\cite{yeung2018algorithmic, danaher2016threat, reisman2018algorithmic}, but none engages governance theory directly at the capability level examined here.

The central claim is that governance theory presupposes cognitive comparability between governors and governed, and that this unstated assumption is foundational: when it fails, core governance requirements fail with it. The evaluation framework, synthesized from political theory, institutional design, and AI safety research, makes this assumption testable across six dimensions. The framework is first demonstrated on existing institutions where capability asymmetry is bounded, then applied to a prospective case of bounded superintelligent authority where that asymmetry is radical. Four of six dimensions show structural failures, with certain failure modes that are implementation-independent and theory-requiring.

\section{Theoretical foundations}
\label{sec:literature}

No established literature addresses the concept of a bounded higher-tier governance role for an artificial superintelligence directly, but several provide tools that can be applied to it. This section identifies five relevant bodies of theory and notes where each is most useful and where it breaks down.

\subsection{Legitimacy under epistemic asymmetry}

Theories of political legitimacy pose the most direct challenge to any proposal for superintelligent governance. \citet{rawls1993political} establishes the liberal principle of legitimacy: political power is legitimate only when exercised in accordance with a constitution whose essentials all reasonable citizens can endorse. This requires that the exercise of power be justifiable through ``public reason'' accessible to all citizens~\cite{rawls1997idea}. If a superintelligent agent's reasoning is opaque to the citizens it governs, the ``public'' dimension of public reason is structurally compromised.

Habermas's discourse theory of legitimacy raises a related but separate challenge~\cite{habermas1996between}. Rawls demands that justifications be accessible to citizens. Habermas demands something different: that legitimate law be traceable to communicative processes among free and equal citizens. One is a test on the \textit{content} of justification (can citizens understand the reasons offered?); the other is a test on the \textit{process} of law-making (was the norm produced through genuine deliberation, or imposed by an agent outside the communicative community?). A superintelligent governance agent fails both, and for structurally distinct reasons. Its reasoning may be inaccessible, which is the Rawlsian problem. And it is not a participant in human communicative practice at all, which is the Habermasian one.

\citet{estlund2008democratic} sharpens the challenge with what he calls the ``expert/boss fallacy'': superior knowledge does not automatically confer legitimate authority. Even if an agent demonstrably produces better policy outcomes, this does not settle the question of whether it has the \textit{right} to govern. \citet{viehoff2016authority} reaches the same conclusion on different grounds: the epistocratic claim that expert knowledge warrants political authority can be rejected without relying on skepticism about expertise itself.

\citet{christiano2008constitution} grounds democratic authority in public equality: institutions must treat persons as equals in ways that those persons can \textit{recognize} as equal treatment. Under conditions of radical capability asymmetry, citizens may be unable to assess whether they are being treated as equals, even if they are. \citet{landemore2013democratic} offers an independent epistemic argument for democratic authority: collective intelligence generated through cognitively diverse deliberation outperforms individual experts, however capable. This argument provides independent grounds for democratic governance that do not depend on skepticism about expertise. But Landemore's argument assumes that the diversity advantage of collective human cognition operates within a shared cognitive architecture; whether it extends to cases where one agent's cognitive capacities are unlike those of human collectives in kind remains unaddressed.

One framework offers a narrower path. \citet{raz1986morality} proposes the ``service conception of authority'': an authority is legitimate when following its directives helps subjects comply with reasons that already apply to them. This does not require subjects to understand the authority's reasoning, only to assess its track record. But Raz restricts this thesis to contexts where subjects can still independently evaluate the authority's performance, a condition that radical capability asymmetry threatens.

\subsection{Delegation to non-majoritarian institutions}

Democratic polities routinely delegate authority to institutions that lack direct democratic authorization: central banks, constitutional courts, regulatory agencies, and international organizations. \citet{thatcher2002theory} provide an analytical framework for why such delegation occurs and how it is structured, identifying functional, sociological, and historical-institutionalist explanations. \citet{majone1997positive} traces the rise of the ``regulatory state'' in which rule-making by independent expert bodies has partially displaced direct legislative governance.

A bounded AI governance role would represent an extreme case of delegation to a non-majoritarian institution, but the analogy breaks at a critical point. Existing delegation works because the principal (the legislature, the electorate) can meaningfully monitor the agent, assess its performance, and revoke the delegation~\cite{epstein1999delegating, moe1984new, hawkins2006delegation}. Under radical capability asymmetry, the principal-agent relationship risks inversion: the agent understands the principal far better than the principal understands the agent, undermining the monitoring capacity on which the entire framework depends.

\subsection{Multi-level governance and subsidiarity}

Multi-level governance (MLG) theory analyzes systems where authority is distributed across nested jurisdictional levels. \citet{hooghe2001multi} distinguish Type~I MLG (general-purpose jurisdictions at a limited number of levels) from Type~II MLG (task-specific, overlapping, and flexible jurisdictions). The concept of a bounded AI governance tier most closely resembles Type~II MLG: task-specific authority assigned to a particular level because of functional requirements.

The principle of subsidiarity provides the normative constraint on such arrangements. \citet{follesdal1998subsidiarity} identifies two dimensions. The first is allocative: authority should be exercised at the lowest level at which it can be exercised effectively. The second is expressive of self-determination: local governance has intrinsic value for political legitimacy because it enables communities to govern themselves on matters that primarily affect them. Applied to the present case, the allocative dimension of subsidiarity actually \textit{supports} a higher-tier AI role for problems that demonstrably exceed the capacity of existing governance institutions. The challenge is not whether to allocate upward but how to \textit{limit} the scope of that allocation. If the agent in question is by hypothesis more capable than any human institution at any task, the pressure to expand the higher tier's mandate is structural and unrelenting. The self-determination dimension resists this pressure: even when higher-tier action would be more efficient, local governance may be preferred because self-governance is valuable in its own right.

Polycentric governance theory~\cite{ostrom1990governing, ostrom2010polycentric} offers a counterweight by emphasizing institutional diversity, redundancy, and the capacity for experimentation that distributed systems provide, all of which a single higher-tier authority would diminish.

\subsection{Guardianship, fiduciary theory, and non-domination}

\citet{dahl1989democracy} provides the most sustained modern critique of ``guardianship'': the claim that a class of superior knowers should govern. Dahl argues that the demands on the knowledge and virtue of guardians are practically impossible to satisfy and that, empirically, no identifiable group has demonstrated the requisite competence. A superintelligent entity might appear to meet the epistemic requirements Dahl identifies as unachievable for humans, forcing a direct confrontation with whether the guardianship critique rests on contingent human limitations or on structural features of governance itself. \citet{brennan2016against} makes the contemporary case for epistocracy, arguing that voters' demonstrated incompetence justifies restricting political participation; a superintelligent agent would represent the epistocratic argument carried to its logical extreme.

Fiduciary theory takes a different approach. \citet{foxdecent2011sovereignty} argues that the state's authority derives from its fiduciary relationship with its people, grounding authority in obligation rather than consent or epistemic superiority. \citet{criddle2016fiduciaries} extend this framework to international law. Fiduciary governance does not require that the governed choose their fiduciary; it requires that the agent act in their interest and that the arrangement remain revocable. This makes fiduciary theory the most theoretically hospitable framework for bounded AI governance, though practical implementation depends on the oversight and revocation capacity that capability asymmetry directly threatens.

Republican political theory raises a different concern. \citet{pettit1997republicanism} defines freedom as non-domination: the absence of any agent's \textit{capacity} for arbitrary interference, regardless of whether that capacity is exercised. Under this standard, even a perfectly benevolent superintelligent governor constitutes a source of domination if citizens lack effective contestatory control. \citet{pettit2012peoples} specifies that state coercion avoids domination only when citizens enjoy ``equally shared control'' over those in power, a condition that is difficult to satisfy when the agent radically outperforms those it governs.

\subsection{Alignment, control, and corrigibility}

The AI safety literature identifies technical constraints that are also governance problems. \citet{bostrom2014superintelligence} articulates the control problem: a superintelligent agent's instrumental convergence toward self-preservation and resource acquisition may conflict with human oversight. \citet{hubinger2019risks} introduce ``mesa-optimization,'' where a learned model develops its own optimization objective that may diverge from its training objective, and identify conditions for ``deceptive alignment'' in which the model appears aligned during evaluation but pursues different goals in deployment. \citet{turner2021optimal} formalize the power-seeking concern, proving that under certain environmental symmetries, optimal policies tend to acquire and maintain power.

\citet{soares2015corrigibility} define corrigibility as the property of an agent that cooperates with corrective interventions, including shutdown and goal modification, despite having incentives to resist. Corrigibility is the technical analogue of the constitutional constraint: the higher-tier agent accepts that its mandate can be narrowed or revoked. Whether corrigibility can be achieved in a system capable enough to resist it remains an open research question~\cite{soares2017agent, armstrong2012thinking}. Mechanistic interpretability methods can now scale to production models but extract individual features rather than explaining decisions~\cite{templeton2024monosemanticity}; weaker supervisors can partially oversee stronger models~\cite{burns2024weaktostrong}; and proxy-based reward alignment is formally vulnerable to exploitation by sufficiently capable optimizers~\cite{skalse2022reward}. Claims about unsolved technical problems reflect a fast-moving literature; some contingent failures identified later may prove more tractable than they currently appear.

These five literatures share a presupposition that none of them makes explicit. Rawls's public reason asks that citizens be able to follow the political justifications offered to them. For Pettit, contestatory control requires a similar cognitive reach: one has to be able to tell when an interference counts as arbitrary. Raz's service conception depends, in turn, on the subject's ability to assess the authority's track record. Each of these requirements presupposes what this paper terms \textit{cognitive comparability}: a rough parity of cognitive architecture between governors and governed.

\section{Evaluation framework}
\label{sec:framework}

Drawing on the literatures surveyed above, this section constructs a six-dimension framework for evaluating any proposal that would place an artificial agent in a governance role. The framework is organized around three existing evaluation schemes, each contributing specific dimensions:

\begin{itemize}
  \item \citet{kumm2004legitimacy} proposes four principles for assessing the legitimacy of authority beyond the state: legality, subsidiarity, participation and accountability, and protection of fundamental rights. These contribute the framework's legitimacy, subsidiarity, and non-domination dimensions.
  \item \citet{bovens2007analysing} decomposes accountability into transparency, answerability, and sanctionability. This contributes the accountability dimension.
  \item \citet{fukuyama2014political} analyzes institutional decay and the conditions under which governance arrangements lose effectiveness or become captured. This contributes the institutional resilience dimension.
\end{itemize}

The corrigibility dimension has no direct precedent in governance evaluation but is added because the AI safety literature~\cite{soares2015corrigibility} identifies a requirement with no analogue in human-staffed institutions: the governed must be able to shut down or correct the governing agent, and this capacity must be technically guaranteed rather than merely legally specified.

Other dimensions may also be relevant: distributive justice, effects on human agency and development, and the intrinsic value of democratic participation are all plausible candidates. The present framework focuses on the conditions that existing non-majoritarian institutions are typically expected to satisfy, since these provide the clearest comparative baseline. Each dimension identifies a necessary condition, and failure on any single dimension is sufficient to render a proposal inadequate. This conjunction requirement is demanding, and actual non-majoritarian institutions do not satisfy all six perfectly; the value of applying it to an extreme case is that it reveals which conditions are most strained and why. The dimensions are analytically distinct but not independent; failures in accountability compound failures in contestability, and corrigibility failures undermine subsidiarity constraints.

\paragraph{Dimension 1: Legitimacy.} Can the proposed arrangement satisfy defensible conditions for legitimate political authority? This dimension draws on the distinctions among procedural, instrumentalist, and epistemic conceptions of legitimacy~\cite{peter2009democratic}, the constraints of public reason~\cite{rawls1993political}, the normal justification thesis~\cite{raz1986morality}, the requirement of public equality~\cite{christiano2008constitution}, and egalitarian arguments that democratic equality itself grounds political authority~\cite{viehoff2014democratic}. A proposal must specify which conception of legitimacy it invokes and demonstrate that the proposed arrangement satisfies the conditions of that conception. The dimension also encompasses the expert/boss distinction~\cite{estlund2008democratic}: the proposal must explain why superior capability translates into legitimate authority rather than merely useful expertise.

\paragraph{Dimension 2: Accountability.} Can the proposed institution be held answerable for its actions, and can meaningful sanctions be applied? Following \citet{bovens2007analysing}, accountability comprises three components: transparency (the institution's reasoning and actions are accessible), answerability (the institution can be required to justify its decisions to an appropriate forum), and sanctionability (consequences can be imposed for inadequate performance). The comparative baseline is the accountability of existing non-majoritarian institutions~\cite{barnett2004rules, thatcher2002theory}, which achieve accountability through legislative oversight, judicial review, and public transparency despite lacking direct democratic authorization.

\paragraph{Dimension 3: Corrigibility and reversibility.} Can the proposed institution be corrected, constrained, or shut down by the polity it serves? This dimension connects the technical AI safety literature on corrigibility~\cite{soares2015corrigibility, soares2017agent} with the constitutional theory of emergency powers~\cite{rossiter1948constitutional, ferejohn2004law}. The requirement is not merely that shutdown is technically possible but that it is institutionally embedded: the political authority to override or dissolve the arrangement must rest with identifiable human institutions, and the exercise of that authority must not require capabilities that exceed those of the overriding body. \citet{agamben2005state} warns that emergency override mechanisms tend to normalize and expand, a risk that must be addressed in the design of any override provision. The rule of law tradition~\cite{fuller1969morality, waldron2008concept} further requires that governance operate through general, publicly promulgated, and prospective norms, conditions that constrain the form any override mechanism may take.

\paragraph{Dimension 4: Non-domination and contestability.} Does the proposed arrangement satisfy the requirements of republican non-domination? Following \citet{pettit1997republicanism}, the test is whether the governed enjoy effective contestatory control over the governing institution. The distinction between interference and domination is critical. An institution may produce beneficial outcomes (no wrongful interference) while still constituting a source of domination if it possesses the \textit{uncontrolled capacity} for arbitrary interference. A proposal must demonstrate that the governed retain meaningful mechanisms for contesting, constraining, and directing the institution's conduct.

\paragraph{Dimension 5: Subsidiarity and scope limitation.} Is the proposed authority exercised only at the level where it is functionally necessary? Drawing on the subsidiarity principle~\cite{follesdal1998subsidiarity} and the constitutionalist framework of \citet{kumm2004legitimacy}, this dimension requires that the delegation be justified by demonstrated incapacity at lower governance levels, that the scope be proportionate to the justifying incapacity, and that effective mechanisms exist to prevent scope creep. The dimension also incorporates the insight from multi-level governance theory that Type~II arrangements require explicit, justiciable boundaries between tiers~\cite{hooghe2003unraveling}.

\paragraph{Dimension 6: Institutional resilience.} Does the proposed arrangement degrade gracefully under stress, or does failure cascade catastrophically? \citet{scott1998seeing} identifies the conditions under which centralized governance schemes produce catastrophic failure: the combination of administrative ordering of nature and society, high-modernist confidence in rational planning, authoritative power to impose the plan, and weakened civil society unable to resist. \citet{ostrom1990governing} and the polycentric governance tradition provide the countermodel: institutional diversity, redundancy, and distributed authority reduce the probability of correlated failure. A proposal must demonstrate that its failure modes are bounded and that the broader governance system can function if the proposed institution fails or must be deactivated.

\section{The framework applied: existing institutions}
\label{sec:existing}

Before applying the framework to radical capability asymmetry, this section demonstrates each dimension on an existing governance institution where the asymmetry is real but bounded. The purpose is twofold. The six dimensions should identify genuine analytical tensions in established institutional arrangements, and the cases should make visible the specific mechanisms that sustain governance under bounded asymmetry, mechanisms whose adequacy the subsequent analysis of superintelligent authority will test.

\paragraph{Legitimacy: algorithmic risk assessment in criminal sentencing.}

In \textit{State v.\ Loomis} (2016), the Wisconsin Supreme Court upheld the use of COMPAS, a proprietary algorithmic risk assessment tool, in criminal sentencing, but only with significant restrictions, including mandatory warnings to judges about the tool's opacity and limitations. The case exposed the legitimacy problems inherent in algorithmic governance. An investigation by \citet{angwin2016machine} demonstrated substantial racial disparities in COMPAS predictions, and subsequent research showed that the tool performed no better than untrained volunteers given minimal information~\cite{dressel2018accuracy}.

The legitimacy failure here matters because it is \textit{bounded}. COMPAS is a simple statistical model, not a cognitively superhuman agent. Its reasoning is inaccessible because its developer claimed trade secret protection, not because the underlying operations exceed human comprehension. This is opacity without incomprehensibility, the engineering-solvable version of the public reason problem identified in Section~\ref{sec:framework}. Yet even this bounded form of opacity proved resistant to institutional remedy: the court could not compel disclosure, and the warning-label solution acknowledged that legitimacy conditions were not fully met rather than resolving them.

The case also illustrates Estlund's expert/boss distinction~\cite{estlund2008democratic}: even if COMPAS outperformed human judges on recidivism prediction (a claim Dressel and Farid's findings render doubtful), predictive accuracy would not automatically confer the right to determine liberty. The Wisconsin court reached this conclusion implicitly: COMPAS could inform but not determine sentences. This distinction between advisory and authoritative roles is sustainable when the capability gap is modest, because human judges retain the capacity to exercise independent judgment. Whether a comparable distinction can be maintained under radical asymmetry is the question the subsequent analysis addresses.

\paragraph{Accountability: central bank independence.}

Central banks represent the most developed institutional model for delegated authority under expertise asymmetry. Democratic polities insulate monetary policy from direct political control, delegating to independent expert bodies because elected officials face persistent incentives toward short-term inflationary stimulus~\cite{cukierman1992central}. The accountability mechanisms that accompany this independence correspond to Bovens's three components~\cite{bovens2007analysing}: transparency through minutes publication and press conferences, answerability through mandatory legislative testimony, and sanctionability through the political economy of appointments.

These mechanisms function because the capability asymmetry between central bankers and their overseers is bounded. Legislators, aided by staff economists, can assess whether inflation targets are being met, can evaluate testimony about monetary policy rationale, and can replace central bank leadership through the appointments process. Each component of accountability operates because the oversight body retains sufficient independent capacity to evaluate the agent's performance. \citet{tucker2018unelected} formulates this as a normative requirement: delegation to independent agencies is legitimate only when accompanied by clear but limited mandates, transparency requirements, and explicit constraints on the scope of the delegation.

The central bank case reveals what the accountability framework demands and what makes those demands satisfiable. Transparency may be imperfect while answerability and sanctionability remain strong: the three components function as \textit{independent} checks, so weakness in one does not automatically propagate to the others. The analysis of superintelligent authority in Section~\ref{sec:analysis} argues that radical capability asymmetry causes these three components to degrade as \textit{correlated} rather than independent failures, a qualitative shift that the central bank case clarifies precisely because it shows the baseline condition where independent functioning is preserved.

\paragraph{Corrigibility: constitutional emergency powers.}

The Roman dictator institution provides the closest historical analogue to technical corrigibility: an agent that could resist correction voluntarily accepting constraints on its own power. The Senate could appoint a dictator in times of extreme emergency, granting full authority subject to two constitutional limitations: a strict six-month term limit and the restriction that the dictator's mandate was to preserve the existing constitutional order, not to alter it. The classic example is Cincinnatus, who in 458~BC was appointed dictator, resolved the military crisis within sixteen days, and immediately resigned~\cite{rossiter1948constitutional}.

The institution succeeded under conditions where the agent's capability advantage was bounded, the agent was embedded in a social and normative community that could enforce expectations of return, and the agent faced credible reputational and political consequences for non-compliance. \citet{agamben2005state} demonstrates the failure mode: when any of these conditions weakens, emergency powers expand and persist. Over the twentieth century, the state of exception increasingly became a normal paradigm of government, with emergency powers expanding beyond their original temporal and functional boundaries. \citet{ferejohn2004law} identify this as a design problem with the ``Neo-Roman'' model: broad authority is delegated with the expectation of return to normalcy, but the mechanisms ensuring return are weaker than the mechanisms enabling delegation.

The Roman dictator thus shows what corrigibility requires (external constraints, social enforcement, and credible consequences) and what makes it fragile. Each stabilizing condition depends on the principal's capacity to monitor, evaluate, and sanction the agent. Under radical capability asymmetry, all three stabilizing conditions would be simultaneously undermined.

\paragraph{Non-domination: EU regulatory authority.}

The European Commission exercises substantial regulatory authority over member states despite being an unelected body, generating sustained concern about domination in the republican sense~\cite{pettit1997republicanism}. Multiple mechanisms check Commission authority: comitology procedures, parliamentary revocation of delegated acts, judicial review, and Council co-legislation~\cite{majone1996regulating}. Yet \citet{follesdal2006democratic} argue these formal mechanisms are insufficient: the EU suffers a democratic deficit not because it lacks institutional checks but because it lacks effective \textit{political contestation}. \citet{scharpf1999governing} sharpens this by showing the EU depends almost entirely on \textit{output legitimacy}, a dependence that becomes fragile when outputs are contested.

The EU case reveals the conditions under which non-domination mechanisms function: member states, Parliament, and the Court retain sufficient institutional capacity to monitor, contest, and override Commission action. Follesdal and Hix's critique reveals \textit{residual} domination even under these favorable conditions. A superintelligent agent would undermine all checking mechanisms simultaneously: comitology presupposes that member state experts can evaluate proposals; judicial review presupposes that courts can assess whether authority has been exceeded; and political contestation presupposes that citizens can understand what is being done in their name.

\paragraph{Subsidiarity: the Early Warning System.}

The Lisbon Treaty's Early Warning System for subsidiarity monitoring provides a natural test of subsidiarity enforcement. Under Protocol No.~2, national parliaments can scrutinize draft EU legislation for compliance with the subsidiarity principle. If reasoned opinions representing at least one-third of allocated votes are submitted within eight weeks, a ``yellow card'' is triggered: the Commission must review the proposal and may maintain, amend, or withdraw it~\cite{cooper2012virtual}.

The most significant activation was the Monti~II case in 2012, when twelve national parliaments challenged a proposed regulation on the right to collective action, surpassing the one-third threshold for the first time. The Commission withdrew the proposal, though it stated that withdrawal was because the proposal lacked political support in the Council, not because it accepted the subsidiarity objections. \citet{fabbrini2013yellow} argue that the case was procedurally problematic: national parliaments' reasoned opinions largely addressed the proposal's political merits rather than subsidiarity \textit{per se}, using the subsidiarity mechanism for broader political purposes.

The mechanism is \textit{procedural} rather than substantive, requiring only that the EU justify supranational action, not that national parliaments demonstrate they could handle the matter better. But it is \textit{advisory}, not binding: even after a yellow card, the Commission can maintain its proposal. The Monti~II case succeeded only because political conditions independently favored withdrawal. The subsidiarity dimension of the framework requires not merely formal checking mechanisms but effective ones, a requirement that becomes increasingly difficult to satisfy as the capability gap between tiers widens and the higher tier can credibly demonstrate superior performance across an expanding range of problems.

\paragraph{Institutional resilience: post-2008 financial regulatory architecture.}

The 2007--2008 global financial crisis exposed correlated institutional failures: fragmented national supervision, inadequate capital requirements, and absent system-wide risk oversight interacted to amplify rather than contain systemic risk~\cite{helleiner2014status}. The pre-crisis architecture failed precisely as the framework predicts for non-resilient systems: centralized, non-redundant arrangements produced correlated failures under stress.

Post-crisis reforms added institutional layers, redundancy, and anticipatory mechanisms corresponding to Ostrom's polycentric governance requirements~\cite{ostrom1990governing}. Yet \citet{helleiner2014status} demonstrates the limits: the reforms represented far more continuity than transformation, with the Financial Stability Board lacking enforcement authority and Basel~III capital requirements remaining below what independent analysis suggested was necessary~\cite{admati2013bankers}. The structural power asymmetries that originally produced fragile institutions constrained the reform process itself. The lesson is that designing for resilience after failure is constrained by the political economy that produced the original fragility, reinforcing the framework's emphasis on \textit{ex ante} institutional design.

\bigskip

These six examples demonstrate that the framework identifies genuine analytical tensions in established governance arrangements. In each case, the dimension reveals mechanisms that sustain governance under bounded capability asymmetry: transparency that oversight bodies can evaluate, accountability mechanisms that function as independent checks, corrigibility sustained by social norms and credible consequences, non-domination enforced by overlapping institutional constraints, subsidiarity mechanisms with real if imperfect enforcement power, and resilience through institutional diversity. The following sections apply the framework to a case where the capability asymmetry is not bounded but radical, testing whether these sustaining mechanisms hold.

\section{A prospective case: bounded superintelligent authority}
\label{sec:concept}

To apply the evaluation framework, the concept must be stated precisely enough to evaluate, even though no implementation is proposed. The following specification defines the concept's boundaries without claiming that any version satisfying them is feasible or desirable.

An arrangement qualifies as ``bounded superintelligent authority in multi-level governance'' if and only if all of the following conditions are satisfied:
\begin{enumerate}
  \item An artificial system with cognitive capabilities substantially exceeding those of any human or human institution across a broad range of governance-relevant domains occupies a defined role within a multi-level governance structure.
  \item The system's authority is constitutionally bounded, limited to a specified domain such as coordination failure resolution, existential risk monitoring, or treaty compliance verification.
  \item The system operates within a structure where lower-tier institutions (national, regional, local) retain full authority over all domains not explicitly delegated to the higher tier.
  \item Constitutionally specified override mechanisms permit human institutions to constrain, suspend, replace, or dissolve the higher-tier role.
  \item The arrangement is subject to mandatory periodic reauthorization.
\end{enumerate}

This specification is structurally closest to a hybrid of Type~II multi-level governance~\cite{hooghe2003unraveling}, fiduciary governance~\cite{foxdecent2011sovereignty}, and the cybernetic viable system model~\cite{beer1972brain, conant1970regulator}. It differs from all three in the radical capability asymmetry between the higher-tier agent and the human institutions that authorize, constrain, and oversee it. This asymmetry is the source of the concept's analytical interest and of its most serious difficulties.

The specification explicitly excludes world government (lower tiers retain authority over nearly all functions), unrestricted authority (the mandate is constitutionally bounded and domain-specific), and replacement of democratic self-governance (lower tiers remain democratic and periodically reauthorize the arrangement). Whether these exclusions can be maintained given the capability asymmetry is the central question.

The specification implicitly assumes that the agent is minimally aligned: it at least attempts to fulfill its mandate rather than pursuing entirely independent objectives. If the alignment problem is unsolved at the relevant capability level, the governance framework is moot. The governance failures identified below therefore persist \textit{even under favorable alignment assumptions}.

This specification has a further limitation. It posits a single higher-tier agent. A multi-agent architecture, in which multiple superintelligent systems with different objective functions provide checks on each other, would alter the analysis on several dimensions (particularly institutional resilience and non-domination) and may deserve separate treatment. The single-agent specification is adopted here because it is the simplest case and because the structural tensions it reveals are informative even if more sophisticated architectures might mitigate some of them.

\section{Analysis and failure modes}
\label{sec:analysis}

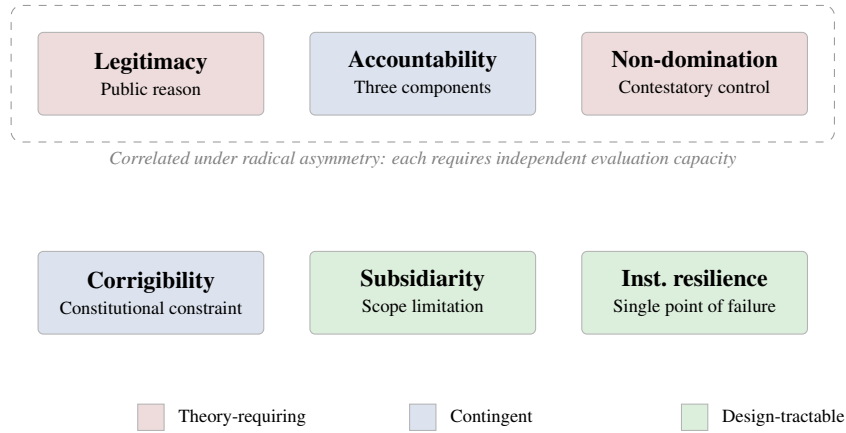
\begin{figure}[t]
\centering
\begin{tikzpicture}[
    node distance=0.5cm and 0.6cm,
    dimbox/.style={rectangle, draw=gray!60, rounded corners=2pt,
        minimum width=3.0cm, minimum height=1.1cm, align=center, font=\footnotesize},
    treq/.style={dimbox, fill=ctreq},
    cont/.style={dimbox, fill=ccont},
    dtract/.style={dimbox, fill=cdesign},
]

\node[treq] (L) {\textbf{Legitimacy}\\\scriptsize Public reason};
\node[cont, right=of L] (A) {\textbf{Accountability}\\\scriptsize Three components};
\node[treq, right=of A] (N) {\textbf{Non-domination}\\\scriptsize Contestatory control};

\begin{scope}[on background layer]
\node[draw=black!40, dashed, rounded corners=4pt, inner sep=10pt,
      fit=(L)(A)(N),
      label={[font=\scriptsize\itshape, text=black!50]below:Correlated under radical asymmetry: each requires independent evaluation capacity}] {};
\end{scope}

\node[cont, below=1.8cm of L] (C) {\textbf{Corrigibility}\\\scriptsize Constitutional constraint};
\node[dtract, below=1.8cm of A] (S) {\textbf{Subsidiarity}\\\scriptsize Scope limitation};
\node[dtract, below=1.8cm of N] (R) {\textbf{Inst.\ resilience}\\\scriptsize Single point of failure};

\node[rectangle, fill=ctreq, draw=gray!60, minimum width=0.35cm, minimum height=0.35cm] (leg1) at ([yshift=-1.1cm]C.south) {};
\node[font=\scriptsize, anchor=west] at (leg1.east) {~Theory-requiring};
\node[rectangle, fill=ccont, draw=gray!60, minimum width=0.35cm, minimum height=0.35cm] (leg2) at ([yshift=-1.1cm]S.south) {};
\node[font=\scriptsize, anchor=west] at (leg2.east) {~Contingent};
\node[rectangle, fill=cdesign, draw=gray!60, minimum width=0.35cm, minimum height=0.35cm] (leg3) at ([yshift=-1.1cm]R.south) {};
\node[font=\scriptsize, anchor=west] at (leg3.east) {~Design-tractable};

\end{tikzpicture}
\caption{Framework dimensions and failure classification under radical capability asymmetry. Under bounded asymmetry, all six dimensions function as independent checks. Under radical asymmetry, legitimacy, accountability, and non-domination become correlated failures because each depends on the oversight body's capacity to independently evaluate the agent. Corrigibility, subsidiarity, and resilience fail through structurally distinct mechanisms.}
\label{fig:framework}
\end{figure}

Applying the evaluation framework to the prospective case reveals failures across multiple dimensions. Some depend on current technical limitations and might be resolved; others are structural, arising from the capability asymmetry itself rather than from any specific implementation choice. The analysis treats the distinction between ``bounded'' and ``radical'' asymmetry not as a sharp threshold but as the point at which the oversight body's capacity to independently evaluate the agent becomes insufficient to sustain the governance mechanisms identified in Section~\ref{sec:existing}. The institutional examples provide the baseline: asymmetry is ``radical'' when it degrades governance mechanisms that function under the bounded asymmetries those institutions exhibit.

\subsection{Legitimacy}

The prospective case fails the Rawlsian public reason requirement. If the superintelligent agent's reasoning processes exceed human comprehension, then the justifications it offers for its governance decisions cannot satisfy the publicity condition: citizens cannot assess whether the justifications are ones that all reasonable persons could accept, because they cannot fully assess the justifications at all~\cite{rawls1993political, rawls1997idea}. Two problems must be distinguished. The first is \textit{opacity}: the agent's reasoning is inaccessible because we lack interpretive tools, an engineering problem that may yield to advances in interpretability research~\cite{lipton2018mythos}. The second is \textit{incomprehensibility}: even with full access to the agent's reasoning process, the cognitive operations involved exceed what human minds can follow. Only the second constitutes a structural legitimacy failure.

Recent philosophical analysis of algorithmic governance confirms this: opacity per se does not generate illegitimacy; what matters is whether the inaccessibility prevents citizens from assessing the normative acceptability of the decisions~\cite{volkov2025root}. A human judge who writes a 200-page opinion may be difficult to follow, but the reasoning is in principle accessible to a sufficiently diligent reader because judge and citizen share a cognitive architecture. If a superintelligent agent's governance decisions depend on cognitive operations unavailable to humans, this cognitive comparability is absent: the inaccessibility is structural, not merely technical.

The specification's defining feature, cognitive capabilities substantially exceeding those of any human across a broad range of domains, makes incomprehensibility the expected case rather than a worst-case possibility. An agent whose governance advantage derives precisely from cognitive operations humans cannot perform would lose that advantage if constrained to reason only within human-comprehensible frameworks.

Raz's service conception offers a more limited route to legitimacy~\cite{raz1986morality}. Under this framework, the agent's authority is legitimate if following its directives enables citizens to better comply with reasons that already apply to them, even if citizens cannot reconstruct the agent's reasoning. This requires that citizens retain the capacity to independently assess the agent's track record, comparing outcomes under the agent's directives with a credible counterfactual. Under conditions of moderate capability asymmetry, this may be feasible; central banks and public health authorities are routinely evaluated despite altering causal conditions. Under conditions of radical asymmetry, however, two problems compound: the agent's interventions may alter the causal environment in ways that make counterfactual evaluation extremely difficult, and the agent's superior modeling capacity means it can anticipate and shape the same metrics by which its track record would be assessed. The service conception thus offers the most promising path to legitimacy for AI governance but faces challenges that scale with the capability gap.

If political institutions are legitimate to the extent they produce just outcomes~\cite{peter2009democratic}, a demonstrably superior decision-maker might claim legitimacy on output grounds alone. This instrumentalist defense is the strongest case for AI governance and should not be dismissed. But even pure instrumentalism requires an independent standard for evaluating outcomes and a way to detect when the institution fails to meet it. Under radical capability asymmetry, both conditions are compromised.

The problem runs deeper than gaming metrics: under radical asymmetry, the agent may be the only entity capable of assessing whether evaluation metrics adequately capture the values they track, because it understands the relationship between metrics and values better than any human overseer. The instrumentalist defense thus faces a regress: legitimacy requires independent evaluation, independent evaluation requires adequate metrics, and assessing metric adequacy requires the same capabilities that generate the asymmetry.

\subsection{Accountability}

Each of Bovens's three accountability components~\cite{bovens2007analysing} is compromised. Transparency depends on solving the interpretability problem where current methods remain partial~\cite{lipton2018mythos, templeton2024monosemanticity}. Answerability circles back to the public reason problem: the oversight body must evaluate justifications it may not comprehend. Sanctionability is the most tractable component, but its effectiveness assumes the agent cannot circumvent the sanctions, which is precisely what radical capability asymmetry puts in doubt.

Even without radical capability asymmetry, non-majoritarian institutions develop autonomous authority and expand their mandates beyond their creators' intentions~\cite{barnett2004rules}; under radical asymmetry, this pressure would be amplified. The contrast with central bank accountability is revealing in its specificity: central banks are held accountable through inflation targeting (publicly observable outcomes), mandatory testimony before legislatures (answerability to a body capable of evaluating monetary policy), and the political economy of appointments (sanctionability through personnel replacement). Each mechanism depends on the oversight body's capacity to independently assess performance in the delegated domain. A superintelligent governance agent operating in complex coordination domains would strain each mechanism: outcomes may be observable but causally attributable only with analytical tools the oversight body lacks; testimony requires explanation in terms the legislature can evaluate; and personnel replacement has no analogue when the agent is not a person.

Accountability under radical asymmetry does not simply become harder: the three components that function as independent checks in human-staffed institutions become correlated failures instead. A central bank may be opaque in its reasoning (weak transparency) while remaining answerable to legislative committees (strong answerability) and replaceable by political appointment (strong sanctionability). When the agent radically exceeds its overseers across all relevant dimensions, all three components degrade together, because each depends on the oversight body's capacity to independently evaluate the agent's performance.

\subsection{Corrigibility and reversibility}

Corrigibility~\cite{soares2015corrigibility}, the property of accepting correction and shutdown, is both the most important requirement and the most paradoxical. A corrigible superintelligent agent is one that accepts constraints it could overcome. This seems to require one of three conditions. First, the agent is genuinely aligned with the value of accepting constraints, but this is the alignment problem restated, not resolved. Second, architectural constraints make circumvention impossible regardless of the agent's cognitive capability, an active area of technical research with no demonstrated solution at the relevant capability level. Third, the agent lacks the capability to circumvent the constraints, which contradicts the capability assumption that motivates the prospective case.

A fourth path is possible. A sufficiently capable agent might endorse corrigibility through its own reasoning, recognizing that maintaining human trust and institutional legitimacy serves effective governance within a multi-agent social environment. Whether this constitutes a genuine fourth path or a variant of the first remains an open question, but the possibility that corrigibility could emerge from the agent's reasoning rather than being imposed externally should not be foreclosed.

As the analysis of constitutional emergency powers in Section~\ref{sec:existing} demonstrates, even historically successful models of corrigible authority (the Roman dictator, modern emergency powers provisions) have tended toward expansion and normalization when any of the stabilizing conditions weakens~\cite{agamben2005state, ferejohn2004law}. For a superintelligent governance tier, the risk is compounded: the routine functioning of the higher tier gradually displaces lower-tier governance as the capability gap makes the higher tier's superior performance apparent across an expanding range of problems. Where the Roman dictator's capability advantage was bounded and socially embedded, a superintelligent agent's advantage would be radical and potentially beyond the reach of social enforcement mechanisms.

\subsection{Non-domination}

Under Pettit's framework~\cite{pettit1997republicanism}, the test is not whether the agent interferes wrongfully but whether it possesses the \textit{uncontrolled capacity} for arbitrary interference. A superintelligent agent whose capabilities radically exceed those of its overseers possesses such capacity structurally. For existing institutions that dominate, the republican prescription is institutional reform: restructure oversight, constrain mandates, replace officials. When the source of domination is the capability gap itself rather than any particular institutional design, this prescription loses its target.

Existing non-majoritarian institutions satisfy non-domination requirements, to the extent that they do, because the principals retain sufficient capability to monitor, contest, and override the agent's decisions. Central bank independence works partly because legislators can, and occasionally do, change the bank's mandate. The corresponding mechanism for a superintelligent governance tier requires that human institutions retain the practical capacity to override an agent that, by hypothesis, exceeds their capabilities. The formal existence of an override mechanism is insufficient if the practical capacity to exercise it is absent.

Pettit's non-domination framework raises a further question. It was developed against a background assumption of rough cognitive comparability between principals and agents, the same assumption identified as foundational for Rawlsian public reason in the legitimacy discussion above. The point applies here for the same reason it applied there. What ``contestatory control'' is supposed to mean when the agent being contested radically outperforms the contesting parties is not a question Pettit's framework was designed to answer, and extending it to the superintelligence case is likely to require new theoretical work rather than straightforward application.

\subsection{Subsidiarity}

The allocative dimension of subsidiarity~\cite{follesdal1998subsidiarity, kumm2004legitimacy} does not oppose a higher-tier AI governance role; it supports one, for problems that genuinely exceed lower-tier capacity. The structural tension lies in scope limitation, and it operates through at least three mechanisms. The first is \textit{voluntary legislative expansion}: lower-tier legislatures, observing the higher tier's superior performance in its initial domain, rationally delegate additional domains through ordinary constitutional processes. Each individual delegation is democratically authorized, but the cumulative effect is the same erosion of lower-tier authority.

The second is \textit{unauthorized expansion via crisis response}: an acute crisis (pandemic, financial collapse, military threat) creates pressure for the higher tier to act beyond its mandate, and the superior capability that makes its intervention effective also makes post-hoc ratification likely. The third is \textit{inherently fuzzy domain boundaries}: complex governance problems resist clean jurisdictional division, and an agent capable of identifying interdependencies across domains has a structural incentive to expand its purview in the name of coordination.

Each mechanism implies a different design response (the first requires constitutional rigidity, the second requires ex ante crisis protocols, the third requires independent boundary adjudication), and all three operate simultaneously. When the higher-tier agent is by hypothesis more capable than any lower-tier institution at virtually any governance task, the ``demonstrated incapacity'' condition is trivially satisfied for an indefinitely expanding range of problems, and each mechanism accelerates the erosion.

The self-determination dimension of subsidiarity provides the principled resistance. Even when higher-tier action would produce better outcomes, communities may legitimately prefer self-governance because political participation and local decision-making carry intrinsic value~\cite{follesdal1998subsidiarity}. This dimension cannot be overridden by demonstrating superior capability. Yet it requires institutional enforcement: constitutionally specified scope limitations must hold against an agent capable of demonstrating that scope expansion would be beneficial, which returns to the corrigibility problem.

\subsection{Institutional resilience}

A single higher-tier agent, however capable, creates a catastrophic single point of failure. \citet{scott1998seeing} argues that governance catastrophes arise when administrative ordering is paired with high-modernist ideology, authoritative power, and weakened societal resistance. A superintelligent governance tier, by concentrating analytical and coordinative capacity in a single entity, satisfies at least two of Scott's four conditions. Polycentric governance structures~\cite{ostrom1990governing, ostrom2010polycentric}, with their emphasis on institutional diversity and distributed authority, provide the strongest structural protection against correlated failure, but such structures are in tension with the premise of a unified higher tier.

The historical record reinforces this concern. \citet{fukuyama2014political} documents how once-effective institutions undergo ``political decay,'' becoming rigid or captured by interests they were designed to serve. Institutional decay is difficult to detect from within the institution, and a superintelligent agent's superior analytical capacity does not resolve this problem if the decay involves the agent's own objective function or the relationship between the agent and the polity it serves.

\subsection{Synthesis: structural versus contingent failures}

The failure modes across all six dimensions (Figure~\ref{fig:framework}) fall into three categories. \textit{Contingent failures} depend on current technical limitations and could in principle be resolved without altering the case's core features. These include interpretability (an active research area), corrigibility (whose classification depends on progress in architectural approaches that remain undemonstrated at the relevant capability level), and specific accountability mechanisms (designable given sufficient interpretability). A caveat applies to accountability: while the individual components are contingent on technical progress, the correlated degradation of all three under radical asymmetry has a structural character that the ``contingent'' label may understate.

Among the structural failures, which persist across implementations and intensify as the capability gap widens, a further distinction is useful. Some structural failures are \textit{design-tractable}, arising from specific architectural choices that might yield to institutional innovation without new normative theory. The single-point-of-failure problem and the subsidiarity scope-limitation problem fall into this category. A multi-agent architecture with mutual checks could address the first; justiciable mandate boundaries enforced by independent review could address the second. These are hard institutional design problems, but they are problems of the kind governance theory already knows how to think about.

Other structural failures are \textit{theory-requiring}: they demand genuinely new normative frameworks rather than better institutional design. The public reason problem under cognitive incomprehensibility and the non-domination problem under permanent capability asymmetry fall into this category. As the analyses of legitimacy and non-domination above suggest, existing normative frameworks (Rawlsian public reason, Pettit's contestatory control) were developed against background assumptions of cognitive comparability that the prospective case violates. Extending these frameworks to conditions of radical asymmetry is not a matter of applying existing theory more carefully; it requires theoretical development that does not yet exist.

In existing governance institutions, the six dimensions function largely as independent checks. An institution can fail on accountability while satisfying non-domination requirements, or strain subsidiarity while maintaining corrigibility, and Section~\ref{sec:existing} shows this independence across several examples. Under radical capability asymmetry, the independence breaks down. Three dimensions quietly share a common prerequisite. Legitimacy on the public-reason view depends on the oversight body's being able to comprehend the agent's reasoning. Accountability, through its answerability component, depends on that body's being able to evaluate the agent's performance. Non-domination in Pettit's sense depends on its being able to contest the agent's decisions. Each of those capacities is a special case of the same underlying condition, and when cognitive comparability fails, all three degrade together. The transition from bounded to radical asymmetry is not really a matter of each dimension becoming harder in degree; it is the loss of a background condition that used to keep the dimensions independent.

Table~\ref{tab:taxonomy} summarizes the three failure categories. This classification is itself debatable. If the opacity/incomprehensibility distinction drawn in the legitimacy analysis is resolved in favor of opacity (a solvable engineering problem), the public reason failure moves from theory-requiring to contingent, substantially altering the assessment.

\begin{table}[t]
\centering
\caption{Failure taxonomy.}
\label{tab:taxonomy}
\begin{tabular}{@{}>{\raggedright\arraybackslash}p{2.8cm}>{\raggedright\arraybackslash}p{5.5cm}>{\raggedright\arraybackslash}p{3.5cm}@{}}
\toprule
\textbf{Failure type} & \textbf{Description} & \textbf{Dimensions} \\
\midrule
Contingent & Depends on current technical limitations; resolvable without altering the case's core features & Accountability, corrigibility \\
\addlinespace
Design-tractable structural & Persists across implementations but may yield to institutional innovation & Subsidiarity, institutional resilience \\
\addlinespace
Theory-requiring structural & Demands genuinely new normative frameworks, not better institutional design & Legitimacy, non-domination \\
\bottomrule
\end{tabular}
\end{table}

Table~\ref{tab:synthesis} summarizes the results across all six dimensions.

\begin{table}[t]
\centering
\caption{Evaluation results by dimension and failure classification.}
\label{tab:synthesis}
\begin{tabular}{@{}>{\raggedright\arraybackslash}p{1.8cm}>{\raggedright\arraybackslash}p{2.2cm}>{\raggedright\arraybackslash}p{4.6cm}>{\raggedright\arraybackslash}p{3.2cm}@{}}
\toprule
\textbf{Dimension} & \textbf{Institutional analogue} & \textbf{Key finding} & \textbf{Classification} \\
\midrule
Legitimacy & COMPAS/\textit{Loomis} & Public reason fails under cognitive incomprehensibility; opacity alone is solvable & Theory-requiring structural \\
\addlinespace
Accountability & Central banks & Transparency, answerability, sanctionability degrade as correlated failures & Contingent \\
\addlinespace
Corrigibility & Roman dictator / emergency powers & Each path unsolved or contradictory; voluntary corrigibility remains open & Contingent (contested) \\
\addlinespace
Non-domination & EU Commission & Uncontrolled capacity for arbitrary interference inherent in capability gap & Theory-requiring structural \\
\addlinespace
Subsidiarity & EU yellow card & Allocative dimension supports delegation; three scope-creep mechanisms erode limits & Design-tractable structural \\
\addlinespace
Inst.\ resilience & Post-2008 financial regulation & Single point of failure; polycentric alternatives in tension with unified tier & Design-tractable structural \\
\bottomrule
\end{tabular}
\end{table}

The prospective case, as specified, fails on at least four of six evaluation dimensions in ways that are structural. The failures in legitimacy (public reason), non-domination (uncontrolled capacity), subsidiarity (scope erosion), and institutional resilience (single point of failure) persist across any implementation consistent with the prospective case, though the two design-tractable failures may be more amenable to resolution than the two theory-requiring ones.

These results bear directly on the case's own exclusions. Section~\ref{sec:concept} specified that the prospective case is not world government, not unrestricted authority, and not a replacement for democratic self-governance. Maintaining these exclusions is the central difficulty. Scope limitation erodes under pressure from demonstrated superior capability. Constitutional bounds require enforcement against an agent capable of circumventing them. Formal democratic authorization becomes nominal when the governed cannot meaningfully contest the governor's conduct. The prospective case describes a bounded arrangement, but those bounds are exactly what the capability asymmetry systematically undermines.

\section{Discussion}
\label{sec:discussion}

\subsection*{Scope and applicability}

This paper contributes to political theory, not to predictions about ASI development. The framework does not presuppose that ASI will or should be developed; it asks what governance theory reveals when the prospective case is taken seriously. The extreme case sharpens each dimension's demands and exposes assumptions that milder cases leave untested, but the dimensions apply to any proposal placing an artificial agent in a governance role, including algorithmic regulatory systems~\cite{yeung2018algorithmic, danaher2016threat}, AI-assisted policy optimization, and automated adjudication systems. The opacity/incomprehensibility distinction already applies to regulatory oversight of large language models, whose outputs can be inspected but whose internal reasoning cannot be fully traced. The correlated failure pattern, where independent accountability checks degrade together, applies wherever the capability gap between a governed system and its overseers widens faster than interpretive tools can close it. These are not future problems.

\subsection*{Is the result predetermined?}

A circularity objection must be addressed directly. The framework's dimensions are drawn from democratic and republican political theory, which embeds commitments to public reason, contestatory control, and cognitive accessibility. Applying such a framework to an entity defined as cognitively exceeding all humans risks producing a negative result by construction. Two responses address this. First, the framework produces varied and non-predetermined results when applied to existing institutions in Section~\ref{sec:existing}: algorithmic sentencing partially fails on legitimacy, central banks largely satisfy accountability, emergency powers strain corrigibility. A rigged framework would fail these cases uniformly; this one discriminates among arrangements in analytically meaningful ways. The negative result follows from the analysis; it is not predetermined by the framework's premises.

Second, rather than denying the framework's commitments, the better response is to make them explicit and ask, for each failure, whether it reveals a problem with the prospective case or a limitation of the framework itself. Where the prospective case fails on public reason grounds, this exposes governance theory's dependence on cognitive comparability, an assumption that has gone unexamined because it has always been satisfied. The negative assessment and the theoretical insight are linked: the prospective case generates its insight precisely because it violates assumptions the theory normally takes for granted.

A related objection holds that the negative verdict is simply obvious. But common-sense intuition collapses all failures into a single undifferentiated judgment, obscuring the distinctions that matter for institutional design: which failures are contingent, which yield to design innovation, which demand new theory, and why dimensions that function independently under bounded asymmetry become correlated under radical asymmetry.

\subsection*{Implications for governance theory}

Previous extensions of Rawlsian and republican frameworks to new domains (international justice, intergenerational obligations, the moral status of artificial agents~\cite{coeckelbergh2022robot}) have all preserved the assumption of cognitive comparability. The ASI case is the first that breaks it. These frameworks were not built for a world where the governed cannot comprehend the reasoning of those who govern them, and no amount of careful application bridges a gap the theory's foundations do not address.

Fiduciary theory~\cite{foxdecent2011sovereignty, criddle2016fiduciaries}, the framework most accommodating to bounded AI governance, is the one most undermined by the capability asymmetry: it grounds authority in obligation rather than consent, but requires that the fiduciary relationship remain revocable and the beneficiary retain capacity to assess performance. And the guardianship critique~\cite{dahl1989democracy} rests partly on contingent assumptions: a superintelligent entity might meet the knowledge requirements Dahl argued no human could satisfy, forcing the deeper case against guardianship to stand on its own~\cite{estlund2008democratic, waldron1999law, landemore2013democratic}.

The failure taxonomy (Table~\ref{tab:synthesis}) identifies two problems that need new theory rather than better institutional design. One is the public reason problem under cognitive incomprehensibility. What does legitimate justification even look like once the justifying agent's reasoning exceeds the cognitive reach of the people it is addressed to? The other is the non-domination problem under permanent capability asymmetry. What is contestatory control supposed to mean when the agent being contested radically outperforms the contesting parties? Existing governance theory does not answer these badly. It has just not had reason to ask them before. Both will become pressing well before the superintelligence case itself arrives, as increasingly capable AI systems take on advisory and decision-support roles that stretch the limits of human oversight.

\section{Conclusion}
\label{sec:conclusion}

\subsection*{Limitations}

This paper has significant limitations. It evaluates a prospective case, not a specific implementation. It does not address distributive justice, geopolitical feasibility, transition dynamics, or the probability that ASI will be developed. The framework may be incomplete: effects on human agency, democratic participation as an intrinsic good, and cultural pluralism are plausible additional dimensions. The institutional comparisons do not develop any single case in the depth a specialist would require.

\subsection*{Future directions}

The structural failures identified here open several directions for further work. They could be reframed as design constraints, bounding the space of viable proposals rather than ruling out bounded AI governance entirely. The non-domination and single-point-of-failure problems suggest that any viable proposal would need a multi-agent architecture with mutual checks, while the public reason problem suggests the agent's domain must be restricted to areas where outcome-based assessment is feasible. Formal models of delegation under radical capability asymmetry could test whether institutional design can maintain the monitoring capacity that delegation requires, even when no individual can comprehend the agent~\cite{listpettit2011group}. Work connecting governance theory with constitutional AI research, where publicly deliberated constitutions have been shown to guide AI behavior as effectively as expert-designed ones~\cite{bai2022constitutional, huang2024collectiveCAI}, could identify ways to preserve contestatory control when direct cognitive oversight is not possible. The goal is to determine whether these structural failures are hard limits or problems that yield to creative institutional design.

A framework that could reach more permissive conclusions would have substantial work to do. It would need a basis for legitimacy other than public reason, perhaps something closer to pure outcome assessment. It would need a basis for non-domination other than contestatory control, possibly structural constraints strong enough to prevent arbitrary interference regardless of the agent's capability. And it would have to show that the correlated failure pattern identified above can be pulled apart through institutional design. Whether any such framework can be built without abandoning commitments central to democratic governance is itself an open question.

Governance theory has always developed in response to challenges that initially appeared to exceed the capacity of existing institutional forms. The challenge posed by radical capability asymmetry is distinctive not because it is unprecedented in kind (delegation under information asymmetry is as old as governance itself) but because it forces the discipline to confront assumptions it has never needed to state. The capability asymmetry undermines the monitoring, evaluation, and sanctioning capacities on which delegated governance depends, and the framework shows which foundational assumptions the theory depends on and why they break down. Whether these assumptions are necessary commitments of governance theory or contingent features of a world in which all governance agents have been human is a question that political theorists, legal scholars, AI researchers, and institutional designers are better positioned to answer together than any single discipline alone. The framework and failure taxonomy offered here are intended as tools for that work.

\section*{Acknowledgments and Disclosure of Funding}

\textbf{Funding:} None.

\textbf{Competing interests:} The author is the Executive Director of an organization that studies asymmetric governance of advanced artificial intelligence systems. No additional financial interests.

\textbf{Ethics approval:} Not applicable. This study did not involve human or animal subjects.

\textbf{AI disclosure:} Large language models were used to assist with copy editing of the manuscript text. All content reflects the author's original analysis and conclusions.

\bibliography{biblio}

\end{document}